\documentclass[%
 reprint,
 amsmath,amssymb,
 aps,
]{revtex4-2}

\usepackage{jabbrv}
\usepackage{xr}

\usepackage{placeins}

\usepackage{graphicx}
\usepackage{bm}
\usepackage{xcolor}

\DeclareSymbolFont{matha}{OML}{txmi}{m}{it}
\DeclareMathSymbol{\varv}{\mathord}{matha}{118}

\begin{document}

\title{Quantifying human performance in chess}

\author{Sandeep Chowdhary}
\affiliation{Department of Network and Data Science, Central European University, 1100 Vienna, Austria}

\author{Iacopo Iacopini}
\affiliation{Department of Network and Data Science, Central European University, 1100 Vienna, Austria}

\author{Federico Battiston}
\affiliation{Department of Network and Data Science, Central European University, 1100 Vienna, Austria}

\begin{abstract}
From sports to science, the recent availability of large-scale data has allowed to gain insights on the drivers of human innovation and success in a variety of domains. Here we quantify human performance in the popular game of chess by leveraging a very large dataset comprising of over 120 million games between almost 1 million players. We find that individuals encounter hot streaks of repeated success, longer for beginners than for expert players, and even longer cold streaks of unsatisfying performance. Skilled players can be distinguished from the others based on their gaming behaviour. Differences appear from the very first moves of the game, with experts tending to specialize and repeat the same openings while beginners explore and diversify more. However, experts experience a broader response repertoire, and display a deeper understanding of different variations within the same line. Over time, the opening diversity of a player tends to decrease, hinting at the development of individual playing styles. Nevertheless, we find that players are often not able to recognize their most successful openings. Overall, our work contributes to quantifying human performance in competitive settings, providing a first large-scale quantitative analysis of individual careers in chess, helping unveil the determinants separating elite from beginner performance.
\end{abstract}

\maketitle

\section{Introduction}
Countless individual careers shoulder the forward momentum in the sciences \cite{sinatra2016quantifying,deville2014career,jia2017quantifying,zeng2019increasing,fortunato2018science}, arts \cite{liu2018hot,fraiberger2018quantifying,Williams2019QuantifyingAP,Janosov2020SuccessAL,Janosov2020ElitesCA} and sports \cite{stevenson1990early,conzelmann2003professional,mallett2004elite,stambulova2013athletes}. Indeed, the recent availability of large-scale datasets is nowadays providing an unprecedented opportunity to study the drivers of human performance in all such different domains. In science, for example, Google Scholar allows to track individual careers of academics, where individual performances can be quantified via their impact, i.e. the attention received from the research community in form of citations.
 In areas like arts, despite the definition of \emph{quality} being somehow elusive, new data has recently shed light on the role of early career exhibitions in reputed venues in the eventual success of the artist \cite{fraiberger2018quantifying}.

 Data-driven investigations of individual activity are a pillar of sports performance. Who can forget the success story described in the book \emph{moneyball} for baseball \cite{lewis2004moneyball}? The day Sandy Alderson realized that on-field strategies and player evaluations were better conducted by based on statistical data—than by the collective wisdom of old baseball men \cite{lewis2004moneyball}, the game---as we knew it---changed. In tennis, network techniques suggested the identification of Jimmy Connors as the best player of the past\cite{radicchi2011best}, a difficult task which requires arbitrary external criteria when comparing across eras. All in all, sport analytics are now commonplace in most major sports, providing clues for individual and team performance to boost success rates \cite{Nevill2008TwentyfiveYO,andrew2019research}. Interestingly, while sports have benefited from scientific methods, they have in turn become a frontier to develop new scientific tools to investigate  success, innovation and learning, as one of the primary domains where growth and success are measurable in a data-driven fashion.

In this work we focus on individual careers in the competitive sport of chess. Moreover, chess is a highly intellectual activity which shares similarities to science. Thus, it is often located amid the two domains, a game where players use simple rules resulting in highly complex plays, often developing different personal styles able to influence long-term success in the game. Besides, the volume of online chess games freely available for analyses (several billions), makes chess a perfect candidate for testing hypothesis involving human performance in competitive settings. So far chess has predominantly been looked at at the level of single games. For example, past research focused on the role of memory in games \cite{schaigorodsky2014memory} and showed that opening popularity follows the well-known Zipf's law~\cite{blasius2009zipf}. However, these analyses did not use individual player-level data, treating games from different players on equal footing\cite{schaigorodsky2014memory,blasius2009zipf}, or focused on a small number of players \cite{arabaci2006investigation}. Indeed, little attention has been devoted to individual careers and their evolution. In particular we ask---{\it what separates skilled players from the rest?} Earlier studies found that the answer is not intelligence \cite{bilalic2007does}, and the role of deliberate practice remains heavily debated \cite{charness2005role,campitelli2011deliberate,hambrick2014deliberate}.
\begin{figure*}[]
\centering
(\textbf{a})\includegraphics[scale=.8,trim={0 0cm 0 0cm},clip]{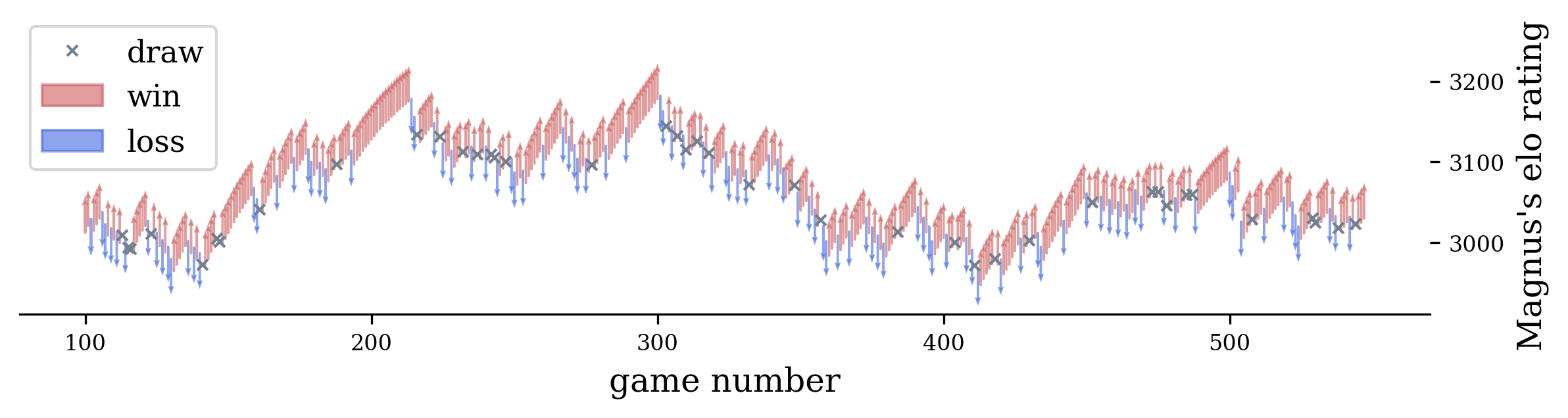}
(\textbf{b})\includegraphics[scale=.8,trim={0 0cm 0 0cm},clip]{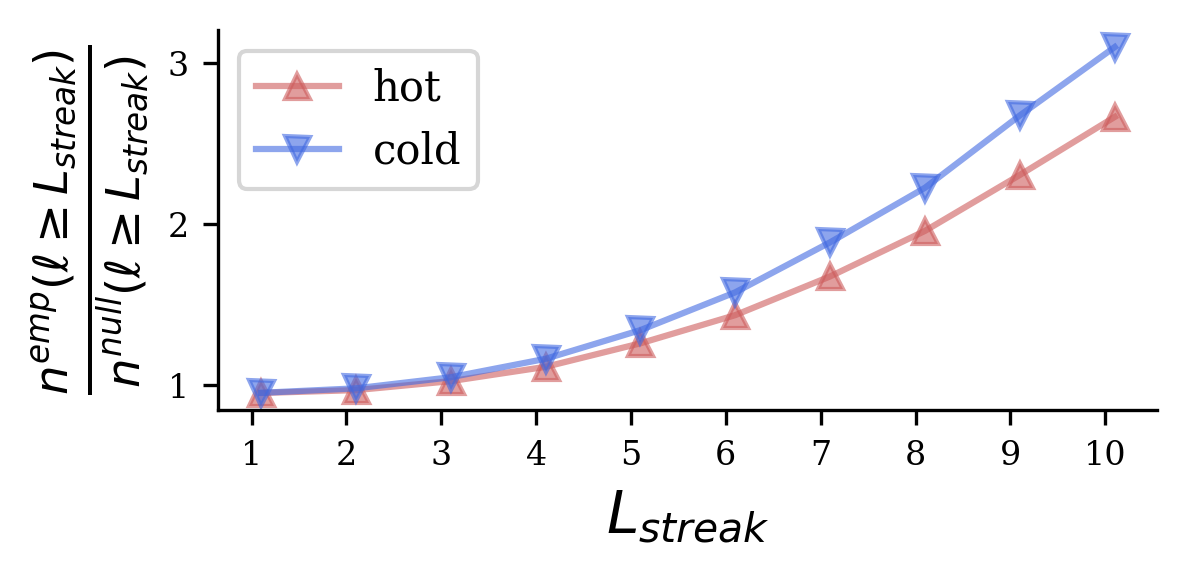}
(\textbf{c})
\includegraphics[scale=.8,trim={0 0cm 0 0cm},clip]{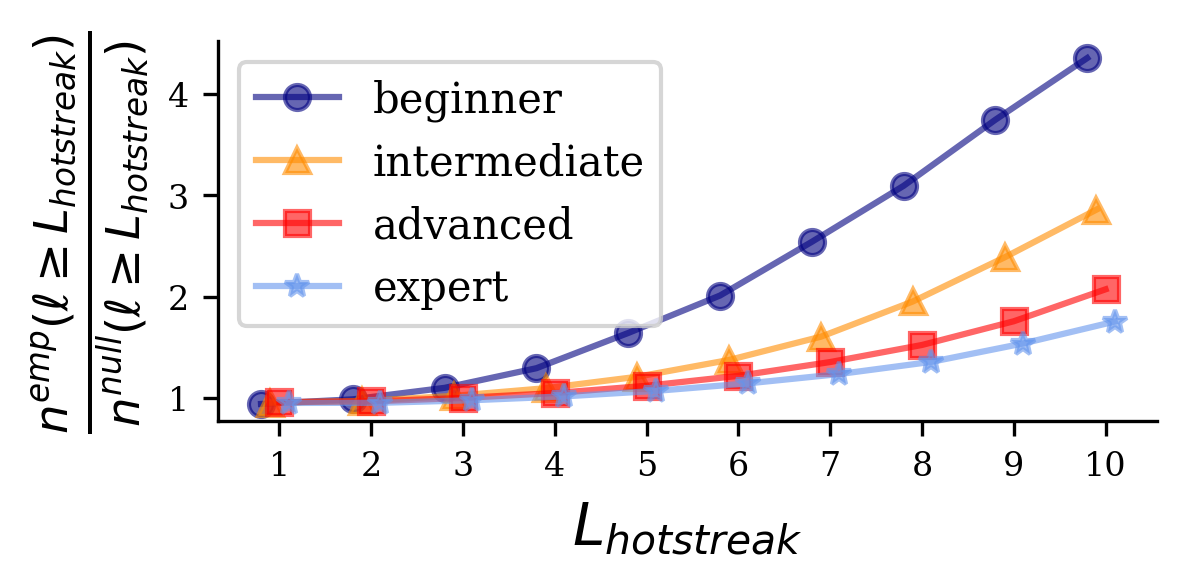}
\caption{\textbf{Hot and cold streaks in chess careers.} Visualization of the career of the Grandmaster (GM) Magnus Carlsen. Wins and losses drive the \emph{Elo} rating up or down. \textbf{(a)} Relative number of hot streaks (red) and cold-streak (blue) of length $\ell \geq L_{streak}$ as a function of $L_{streak}$ calculated for each player. Results are averaged over all players. Losses tend to be more clustered than victories as individual cold streaks are on average longer than hot streaks. \textbf{(b)} Relative number of hot streaks of length $\ell \geq L_{hotstreak}$ as a function of $L_{hotstreak}$, averaged over the players in each skill categories separately (i.e. beginner, intermediate, advanced, expert). Weaker players have longer hot streaks than more expert ones.}
\label{fig1}
\end{figure*} 
Here we perform a comprehensive large-scale analysis of the habits of skilled and less skilled individual players over time, providing an anatomy of human performance in the popular game of chess. We characterize players' careers in terms of hot-streaks, diversity and specialization in the opening sequences of their games, and analyze their diversity as a function of career stage. We find evidence for the presence of both hot and cold streak phenomena, revealing a surprising tendency for beginners to have longer hot-streaks as compared to expert players. By sequencing the opening moves of players at different skill levels, we show that beginners start with more diverse set of first moves, while advanced players and experts rarely start their games differently when playing as white. Yet, expert players display a broader response repertoire, showing the ability to surprise their opponent with a greater variety of responses. Moreover, when accounting for different variations of the openings, experts show a deeper knowledge of different variations within the same line, hinting at a deeper understanding of the game. Lastly, analyzing behaviour in time, we find that players explore more during the beginning of their careers, but tend to specialize using and exploiting only fewer openings at later career stages. Overall, our large-scale characterization of individual gaming behavior supports chess as a suitable laboratory to quantitatively investigate individual careers and human performance, demonstrating simple differences in playing habits and behaviours of beginners and experts.

\section{Results}
In this work, we rely on large-scale data extracted from \emph{lichess.org}, a popular open-source Internet chess server, consisting of 123 million games between 0.98 million players (see Sec.~\ref{data}). In the \emph{lichess} dataset, each player's career can be tracked over time, with detailed information on each of the played games, i.e. moves, opening, win/loss, and its skill level. This is quantified by the \emph{Elo} rating (see Sec.~\ref{elo}), which measures the level of past performance of the player, it increases when a player beats an opponent and decreases upon a loss. As an illustrative example, in Fig.~\ref{fig1}a we show the career of Grandmaster (GM) Magnus Carlsen on \emph{lichess.org}, indicating his \emph{Elo} in each game and the game outcome: win, loss or draw.

Figure~\ref{fig1}a seems to suggest that for \emph{GM Carlsen} wins and losses tend to be clustered together.
Indeed, prior works tracking wins and losses in sports hotly debate the existence of hot-streaks \cite{gilovich1985hot,miller2018surprised}, a phenomenon that has also been found to be ubiquitous in artistic and scientific careers \cite{liu2018hot,liu2021understanding}.

\begin{figure*}[]
\centering
\begin{minipage}{.48\textwidth}
{(\textbf{a}) \includegraphics[scale=.44]{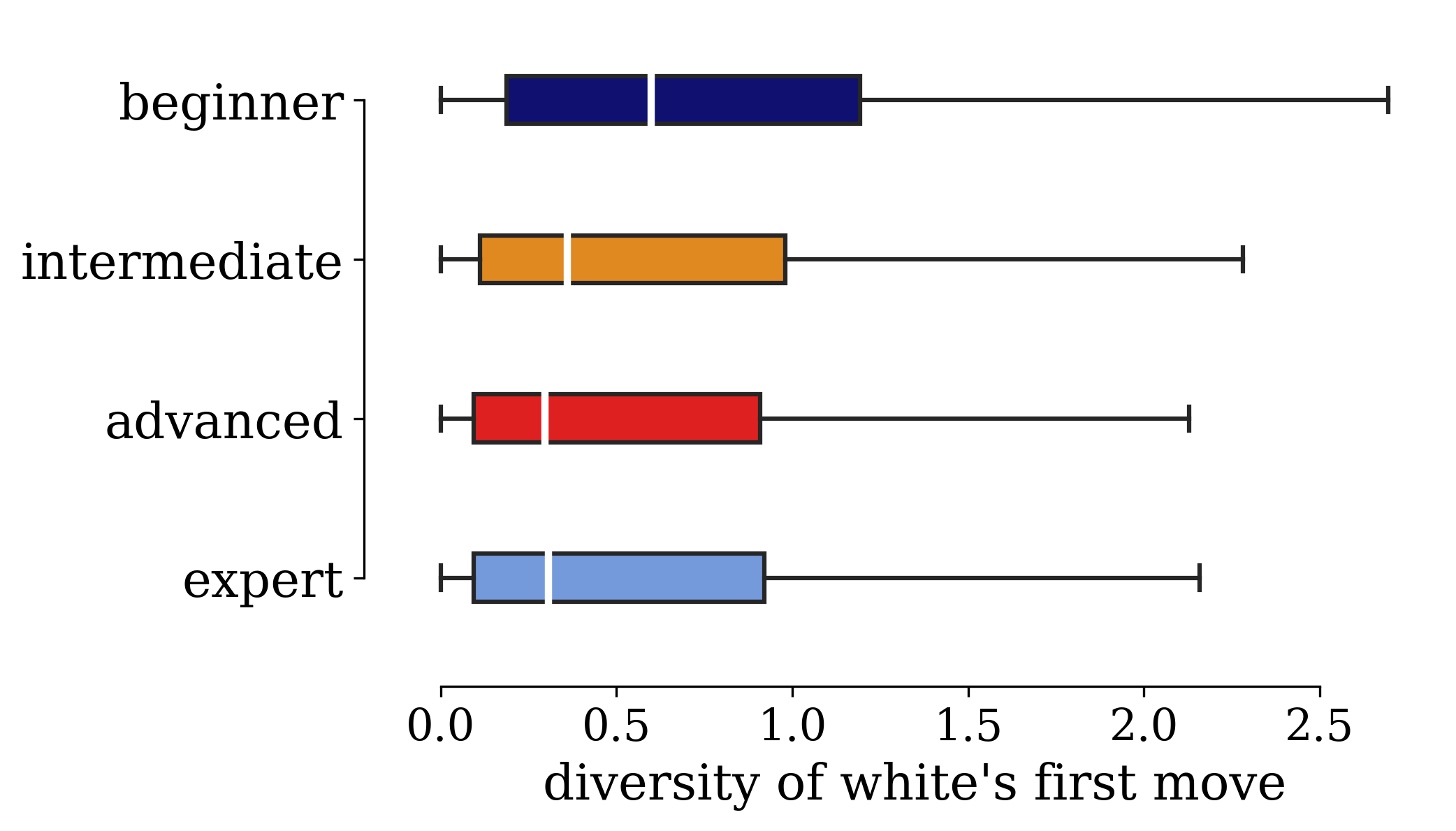}}    \end{minipage}%
\begin{minipage}{.48\textwidth}
{(\textbf{b}) \includegraphics[scale=.45]{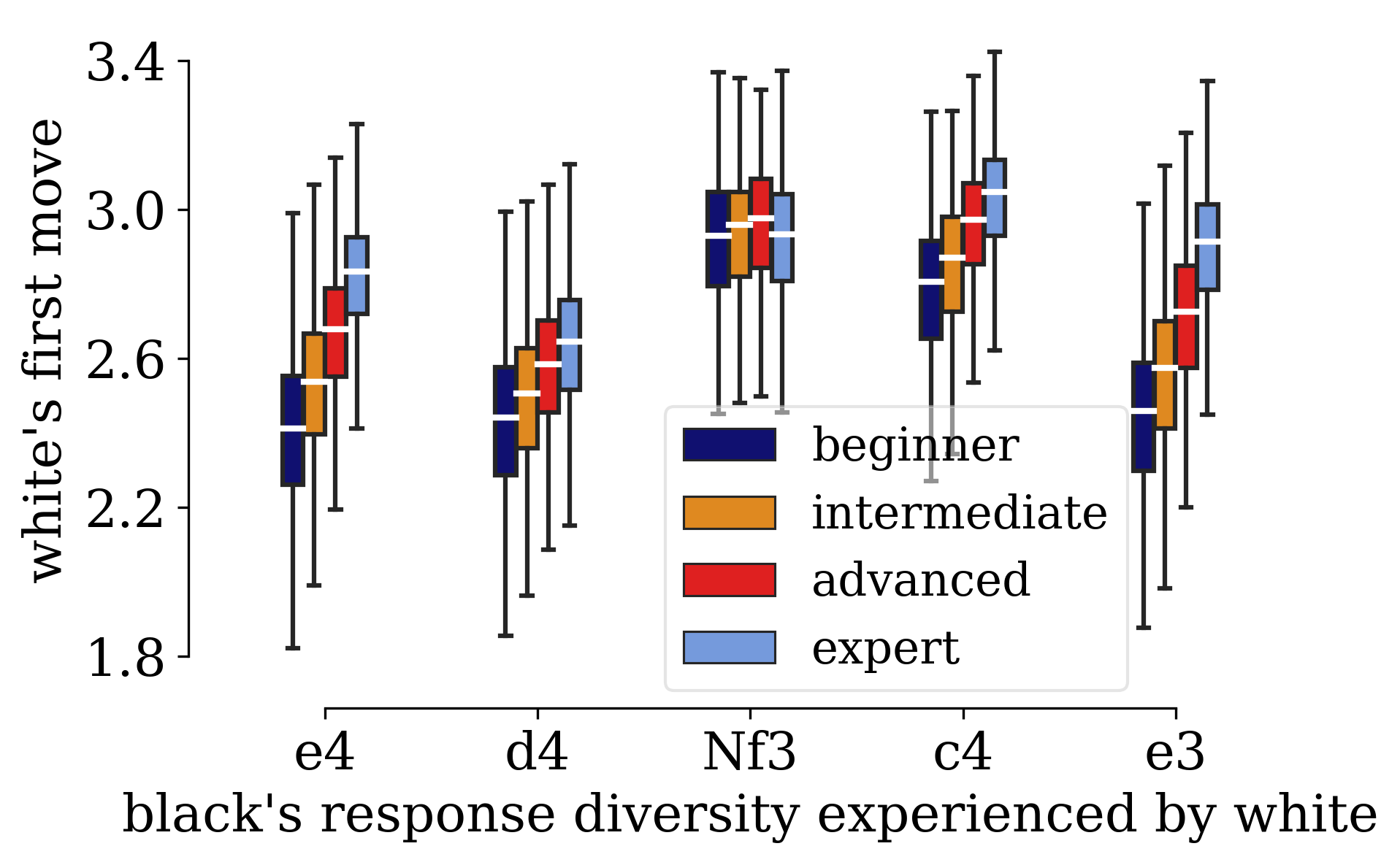}}    \end{minipage}%
   
\caption{\textbf{Diversity  and specialization in the first move and black's response.} \textbf{(a)} Boxplots showing diversity (entropy) of \emph{first move} by a player as white, calculated over all players individually and aggregated into the 4 different skill levels. Weak players start games with diverse collection of first move as white when compared to stronger players. \textbf{(b)}  Boxplots showing diversity of black's response experienced by white player, for each of white's top 5 most played first moves- $e4$, $d4$, $Nf3$, $c4$ and $e3$  (in descending order of popularity). As white, weakest players encounter lowest diversity in responses captured by low response entropy-- for all of white's most played opening moves, except Nf3.}
\label{fig2}

\end{figure*}
\begin{figure*}[!ht]
\begin{minipage}{.43\textwidth}{(\textbf{a})}\includegraphics[scale=.65,trim={0 0cm 0 0cm},clip]{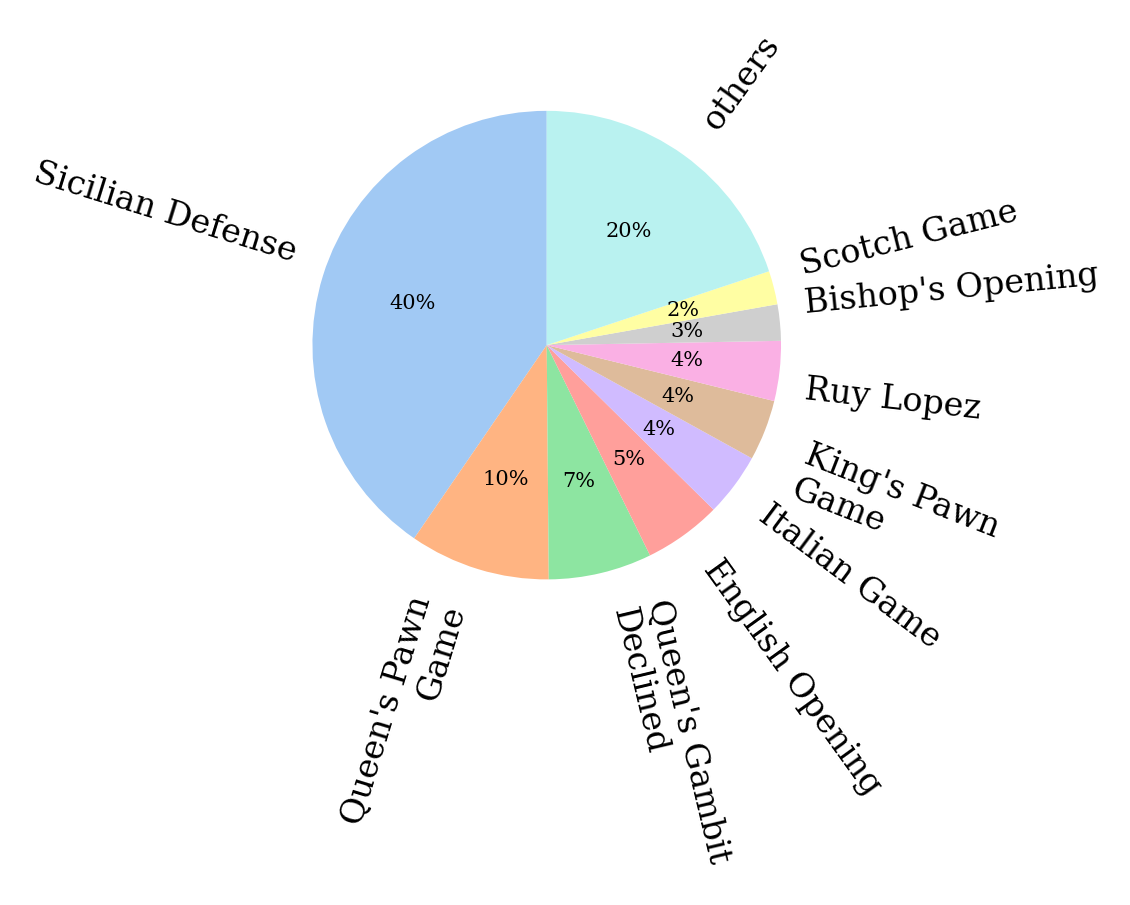}
\end{minipage}%
\begin{minipage}{.55\textwidth}{(\textbf{b})}\includegraphics[scale=.7,trim={0 0cm 0 0cm},clip]{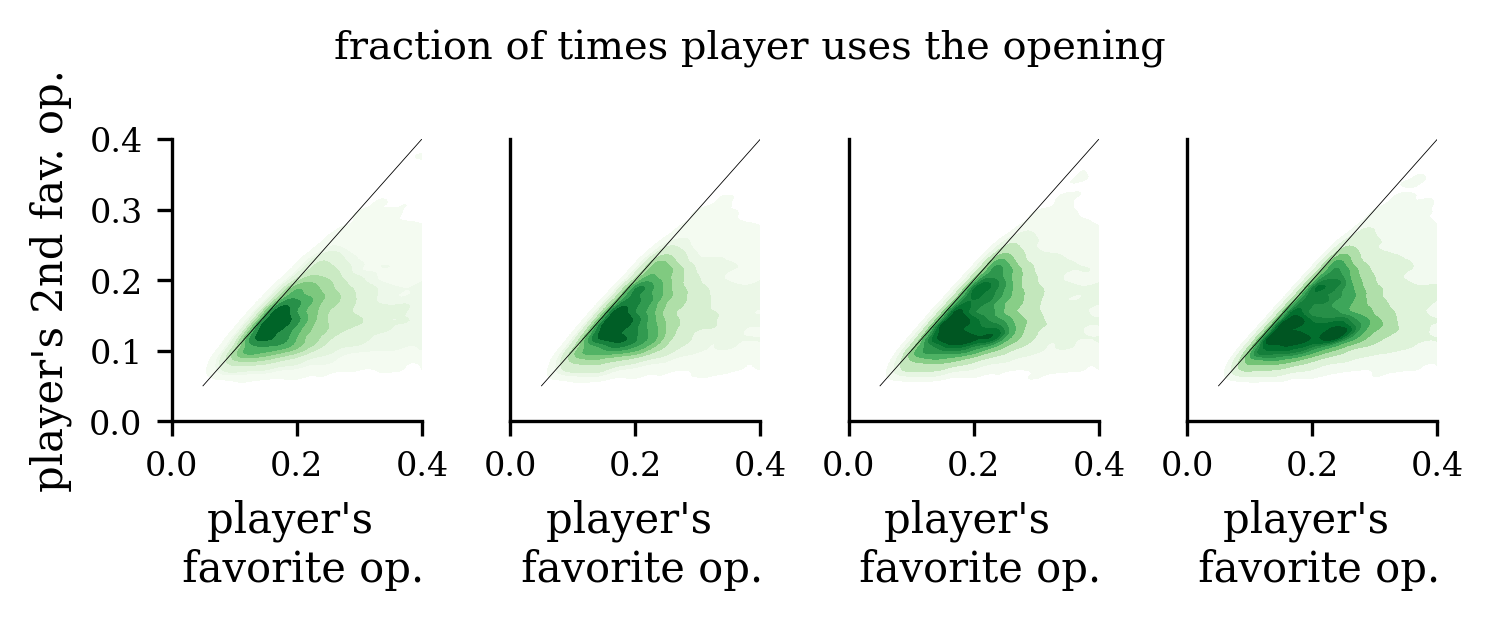}
\end{minipage}%

{(\textbf{c})}\includegraphics[scale=.7,trim={0 0cm 0 0cm},clip]{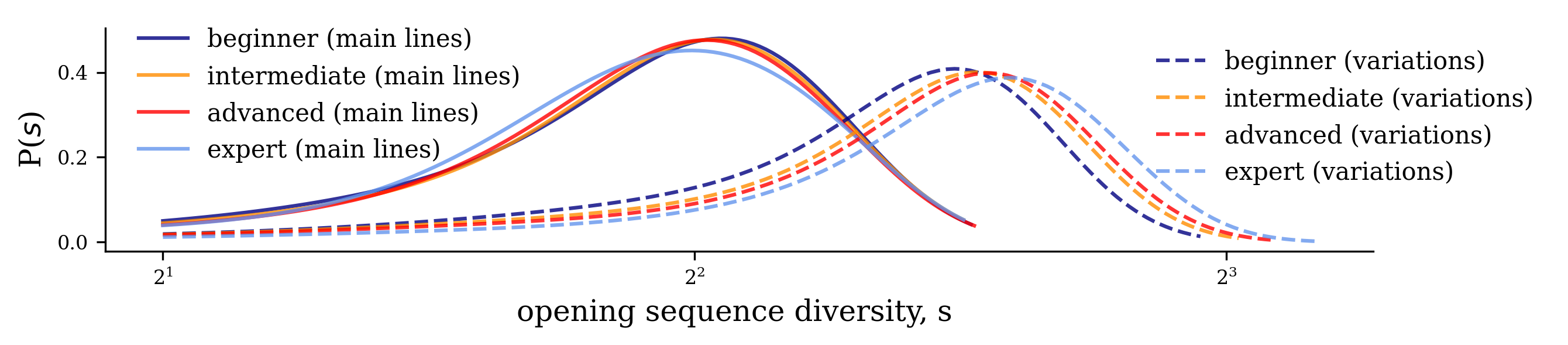}

\caption{ \textbf{Diversity and specialization in the opening sequence of moves is governed by skill level.} \textbf{(a)} Top 9 favorite openings among all players.  \textbf{(b)} Fraction of times players use their top-2 favourite openings. Different panels correspond to different skill levels. Density plots are used for better visualization. Expert players play more often their favorite opening sequence as compared to beginners. \textbf{(c)} Distribution of diversity (entropy) of openings calculated for players of four different skill levels. Main lines and the variations are respectively depicted as solid and dashed curves. }
\label{fig3}
\end{figure*}

To quantitatively check for the existence of such phenomena in all chess careers, we calculate the length of hot (series of wins) and cold (series of losses) streaks for each player in the dataset, and compare them with lengths expected in a reshuffled null model (see Sec.~\ref{null_hot}) for each player. In Fig.~\ref{fig1}b we show the resulting curves, properly normalised with the null model. We find the existence of statistically significant hot streaks, possibly associated with confidence spillovers from previous victories. Long streaks of chess wins are reminiscent of players entering the so-called \emph{zone}, a state of focus where peak performance is possible \cite{young1999zone,murphy2011zone}.

Interestingly, cold streaks are also observed, typically longer than hot ones, indicating that times of poor performance tend to last more than periods of intense success. In physical sports, loosing streaks are often found to be the effect of an injury. Here, we speculate that similar phenomena might be in place even in chess, possibly due to lack of confidence, loss of focus and similar decrease in mind fitness more than purely athletical shape.

We can refine such analysis by further separating players by skill (Elo). Categorizing players into 4 categories - \emph{beginner, intermediate, advanced, expert}~ (see Methods). We find that weaker players experience comparatively longer hot streaks than stronger players (Fig.~\ref{fig1}c). A reason for this could be that confidence spillovers from last victory may have greater impact on future outcomes at a lower skill levels.

Another possible driver of the observed disparity in hot streaks across beginners and experts can reside in how experts diversify their moves. In competitive sports, some players diversify their techniques while others may specialize. Strategy diversification might make players harder to predict, thus enabling them to surprise their opponents. By contrast, specialization, e.g., deeper knowledge of certain opening positions, may allow players to exploit opponents navigating familiar situations. Indeed, such an exploitation-exploration (specialization-diversification) dichotomy is a common mechanism governing the dynamics of many diverse self-organized and adaptive systems \cite{kuhn2011essential, march1991exploration, pappalardo2015returners, iacopini2018network}. 
\begin{figure*}[]
\centering

\begin{minipage}{.45\textwidth}
(\textbf{a})
\includegraphics[scale=.7,trim={.2cm 0cm 0 0cm},clip]{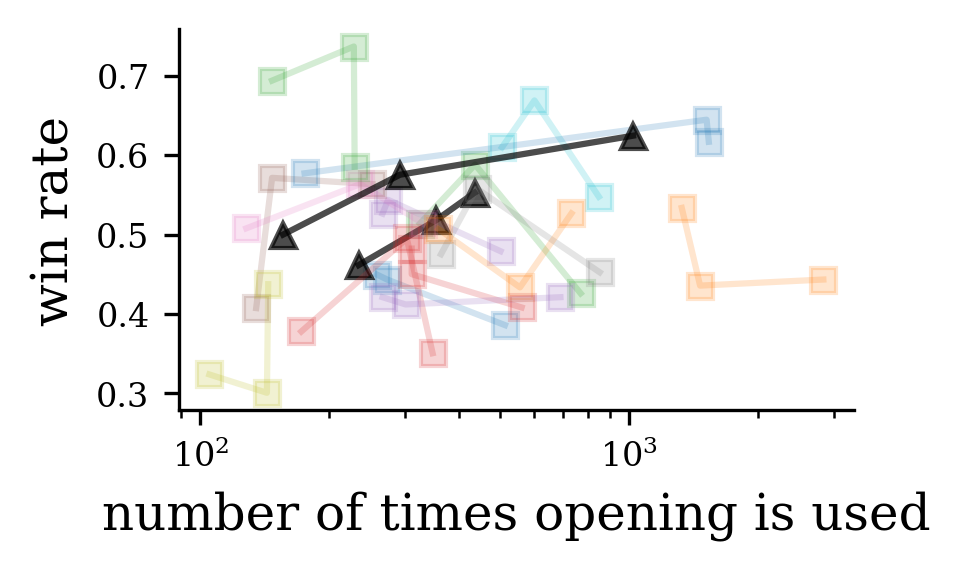}
\end{minipage}
\begin{minipage}{.54\textwidth}
        
(\textbf{b}) \includegraphics[scale=.7,trim={.2cm 0cm 0 0cm},clip]{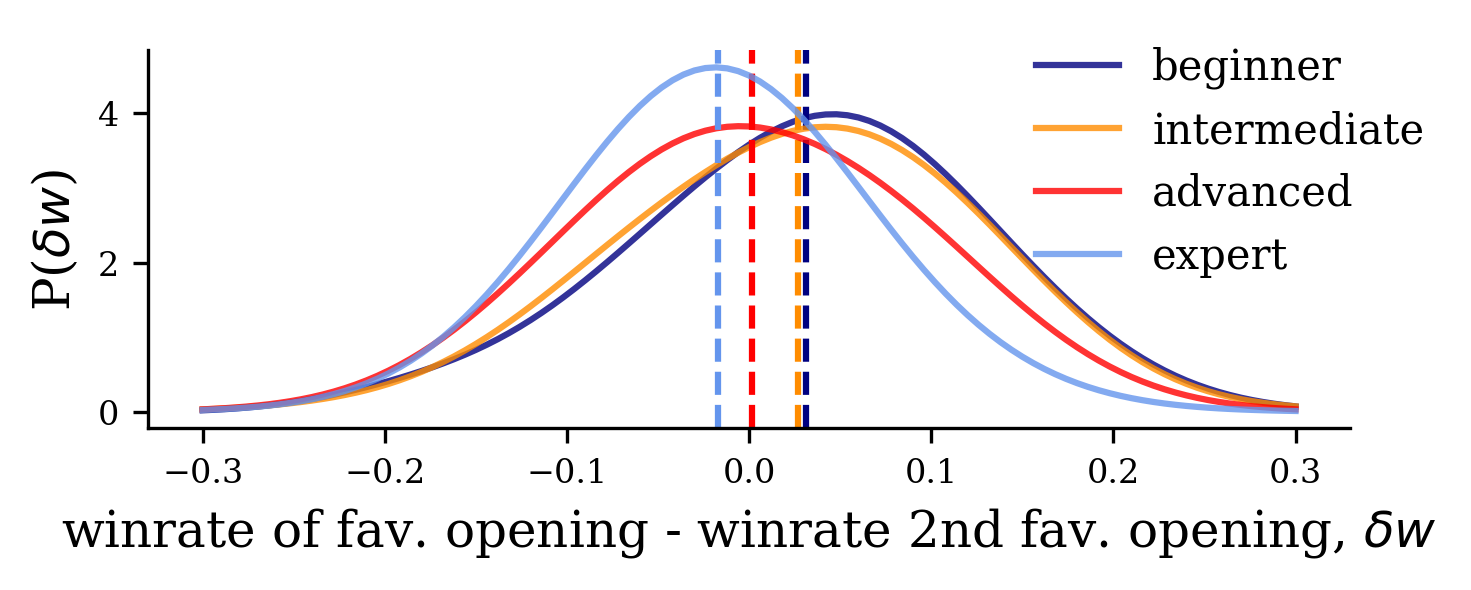}
\end{minipage}
\caption{\textbf{Sub-optimal opening encounters.} \textbf{(a)} Winrate of top 3 openings of a player against opening frequency. Each connected curve corresponds to a player.  We show 15 random players who play at least 100 games with each of their top 3 openings. Curves of players whose winrate increases monotonically with the frequency of the associated opening are depicted in black and are deemed \emph{optimal}. \textbf{(b)} Distribution of difference $\delta w$ in winrate of associated to favourite and second favourite opening and winrate of their second favourite opening for the whole population of players. Different curves correspond to different skill levels. Dashed lines indicate mean values of the distributions. Stronger players encounter less optimal openings more often than weaker players. }
\label{fig4}

\end{figure*} 

In chess---and sports in general---the balance of this trade-off may depend on skills. We thus investigate the extent to which skill level influences the approach to the game. In particular, we study the diversity in the player's arsenal of game openings across different Elo ratings (Fig~\ref{fig2}). We calculate the Shannon entropy of the distribution (see {\it Methods}) of first move as white for each player and report the results in Fig~\ref{fig2}a. We find that beginners tend to open games with a diverse collection of first moves (as white) when compared to stronger players. Thus, our analysis captures beginners exploring a wider variety of first moves than experts, who instead are likely to begin with a typical move. At a first glance, this result might seem surprising, as skilled players are supposed to have better knowledge of opening theory. Yet, this may be linked to the ability of more skilled players to easily transpose into different opening variations in the following moves. Better awareness of transposition theory among experts may allow them to reach many different openings from the same starting move, thus potentially eliminating the need to diversify in the first move itself. 

So, overall, do experts specialize at the cost of diversity? To investigate further, we ask---how does skill level determine response diversity (as black)? For the top 5 white moves observed---$e4$, $d4$, $Nf3$, $c4$ and $e3$, we group the games of each player based on these moves and calculate the response diversity of the black to the white player. Results are shown in the different boxplots of Fig \ref{fig2}b. Surprisingly, we observe a contrasting result. As white, beginners encounter the lowest diversity in black responses. This is captured by the low response entropy for all 5 of white's most played opening moves. Hence, beginners lack experience to the plethora of possible responses, which perhaps leaves gaps in their game.

Lastly, we point out that this increase in the diversity of responses at higher skill levels, might be what prevents players from increasing their Elo, as the potential to be surprised by your opponent keeps increasing as one climbs the Elo ladder.

From the first move onward, players enter into established chess theory, where the many top variations of opening moves are well-explored. The next natural question to ask at this point is---How do players diversify beyond the first move as player move into opening theory? The beginning usually plays out like a well-choreographed dance, evolving in already classified opening sequences with standard names such as ``Sicilian Defense'', ``Queen's Pawn Game'', and so on. In Fig.~\ref{fig3}a, we show the top 9 openings used by players on \emph{lichess.com}. Focusing on such opening sequences, we explore the specialization players achieve in the opening sequence. Results are shown in Fig.~\ref{fig3}b, where we define the "favorite opening" of a player as the most used one, assuming it is played at least 100 times. 

Interestingly, the majority of players end up in their favorite openings only around ~$10\%$ to $30 \%$ of the time. Furthermore, we find that expert players start with their favorite opening significantly more times than their second favourite. This is marked by the distribution falling below the diagonal line.  Contrarily, beginners lie much closer to the diagonal, indicating that their favorite opening is played comparably to the runner up, thus pointing out a lack of opening specialization. 

Further analyses reveal that expert playing behavior comes in a variety of shapes and sizes, i.e., there are players who specialize and players who flexibly switch openings (diversify). This is marked by the bi-modal nature of the distributions observed in Fig.~\ref{fig3}b, column 4. 

At the individual level, we find on average less diversity in opening selection (main lines) among experts, as shown in Fig.~\ref{fig3}c. As mentioned earlier, the ability to arrive into known openings through \emph{transposition}, i.e., different sequences of moves that players may use to reach the same final configuration, might be unique to expert players. Arriving into fewer openings may allow experts to use learned chess theory and use optimal moves from memory, saving crucial time and preventing build-up of mental fatigue during the game. 

However, accounting for the many different ~\emph{variations} of the openings (see {\it Methods}), it is the experts instead who encounter the most diversity. This hints that experts like to enter into certain main openings---perhaps the ones they specialize in---which they follow-up by expanding their repertoire in the \textit{variations} to surprise opponents and catch them off-guard, a strategy not unique to chess but key in many competitive sports. Furthermore, upon investigating temporal organization of openings (main lines) used by a player, we find that experts switch openings between consecutive games more often than beginners (see Supplementary Information, Sec.~S3). Thus, experts encounter higher temporal diversity in openings.

At this point one might wonder---how much exactly does specialized knowledge of favorite openings aid in victory? A naive argument would suggest that players would tend to prefer those main lines that give them the best results. If this is the case, the favourite opening of each player---the one mostly used---would be the one that gives the best performance, that is the highest winrate. To investigate this, we calculate for each player the winrate of each of the player's top-3 most played openings and plot it against the frequency of their use. Results are shown in Fig.~\ref{fig4}a for a sample of the players.  Surprisingly, there are players whose top used opening performs worse than their lesser used openings. Besides, optimal players (black curves)---those who play more often their better performing openings---are just a few.

To quantify this effect in the whole population, we calculate for each player the difference in the winrate of the most played opening and the second most played one, showing its distribution in Fig.~\ref{fig3}b. Our analysis reveals that when expert players do encounter their favorite opening, their winrate is more likely to be lower than their second favourite opening, when compared to beginners. We note that players who do better in their second most played opening---as compared to their most played one---are experiencing sub-optimal opening encounters. Thus, we find that stronger players encounter sub-optimal openings more often than weaker players. Discovered sub-optimal encounters may be an opportunity for players to improve. 

\begin{figure}[!ht]
\centering
(\textbf{a})\includegraphics[scale=.43,trim={.2cm 0cm 0 0cm},clip]{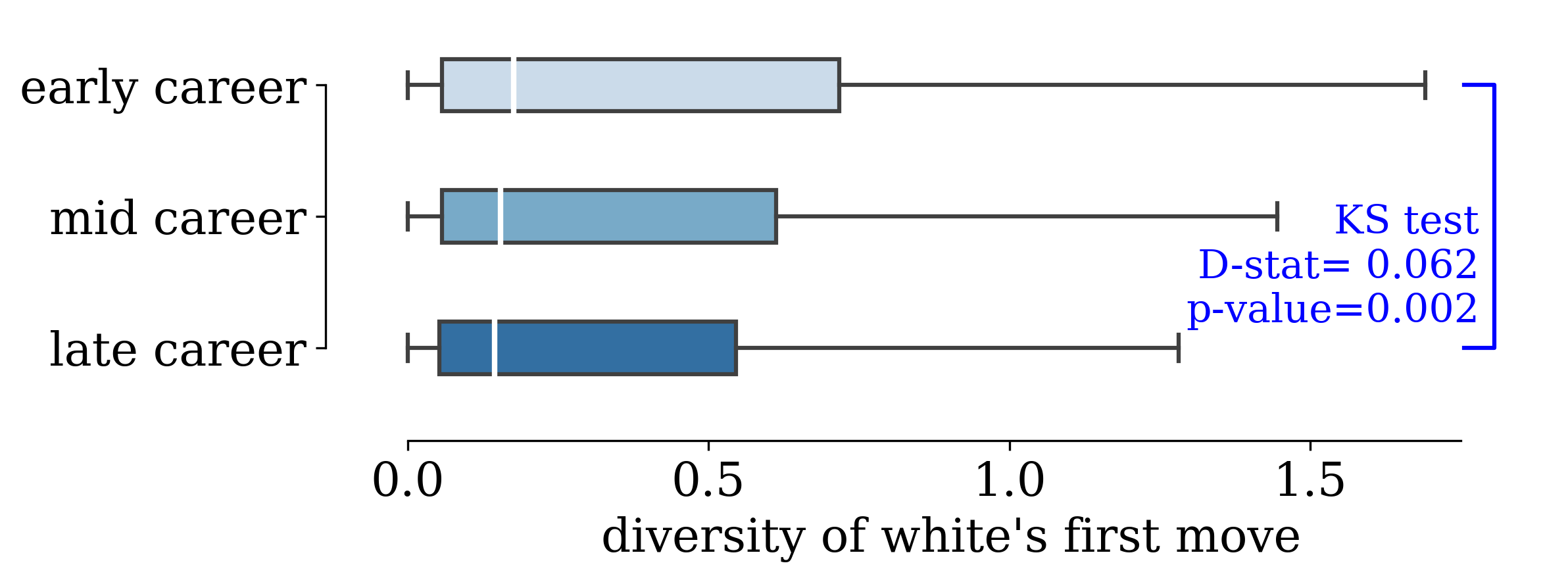}
(\textbf{b})\includegraphics[scale=.43,trim={.2cm 0cm 0 0cm},clip]{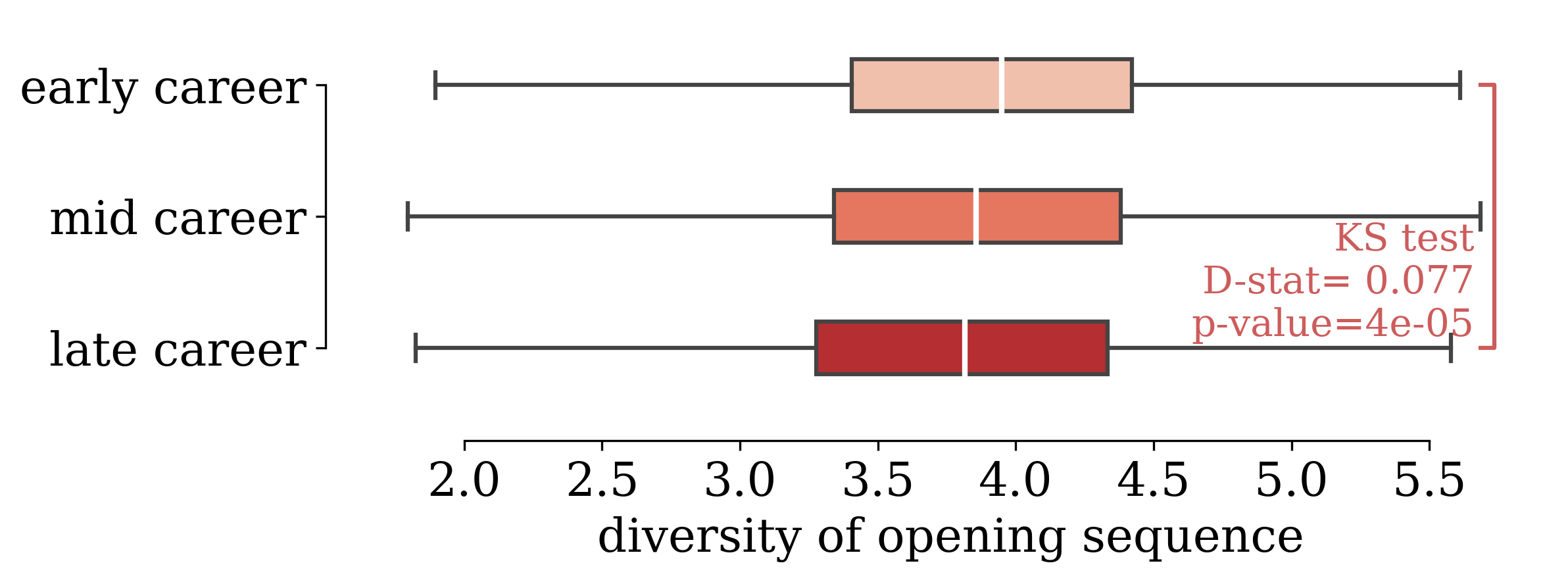}

\caption{\textbf{Diversity in opening with career stage.} We show the diversity in the opening move \textbf{(a)} and opening sequence \textbf{(b)} of moves. In later parts of one's career, diversity decreases, players prefer certain openings and specialize more in them---playing them more often.}
\label{fig5}

\end{figure}

Lastly, we explore diversity as a function of different stages of players' careers. Selecting players with at least 3000 games, we split them into 3 equal stages: early (0-1k), mid (1k-2k) and late career (2k-3k). For each play, we compute opening diversity in the different career stages and report it in Fig.~\ref{fig5}. For both the opening move (a) and the opening sequence (main lines) (b), we find that players explore more in the initial stages of their careers, becoming more specialized in later stages, perhaps exploiting the knowledge of certain openings they have learned.

\section{Discussion}
Quantifying performance and unveiling the drivers of success is an ubiquitous pursuit in modern society,  especially important in competitive settings, where skills, techniques, strategies, and achievements need to be compared. Indeed, in many sports, measuring and analysing performance has nowadays become a common practise \cite{ericsson1996expert,Ericsson1996TheRT}. In this work, we propose chess as a natural laboratory to investigate human behavior  \cite{Groot1978ThoughtAC,simon1988skill,Chase1973PerceptionIC,Simon1973ASO}. Differently from most other disciplines, chess has no stochastic component, hence performance can be more directly associated with skill---as quantified by Elo. Having this in mind, here we performed a data-driven investigation of almost 1 million individual careers carving their way to success in the online platform \emph{lichess.org}. Looking at entire careers, we found the presence of hot and cold streaks. These are bursts of victories and losses, already observed in science and other domains~\cite{liu2018hot,Janosov2020SuccessAL,liu2021understanding}, which might be due to periods of particularly successful physical fitness, creativity or even confidence.
Accounting for skill level, we noticed that beginners have a higher chance to experience a repeated sequence of wins, and that the typical length of a hot-streak is inherently related to the skill level of a player. Moreover, no matter individual ability, player performance is characterized by even longer period of repeated failure. 

Even just looking at simple patterns in the openings---thus neglecting the full complexity of game sequences---, we were able to characterise individual playing behavior across different career stages. In particular, expert players were shown to behave differently from the very first move of the game, displaying a lower diversity in openings. Looking at chess as a process of interactions and reactions, we focused on the black's response to the white player's moves, finding that experts encounter the highest diversity from black. However, after accounting for different variations within the openings we discovered that experts were more diverse instead, hinting at a deeper understanding of the complexity of the different variations within the same line. Such findings corroborates some very recent ideas on opening similarity and complexity independently presented in Ref.~\cite{de2022quantifying}, focusing on prediction of future openings and opening preparation, about which we recently became aware. Looking at individual careers over time, opening diversity was found to decreases at their later stages, pointing towards higher specialization as a player becomes more experienced. In addition, experts tend to play their favorite opening sequence much more than beginners, providing evidence for a tendency towards specialization. Nevertheless, counter-intuitively, we also found that players often do not have the ability to recognize their most successful opening, i.e. the one associated with the highest win-rate. Surprisingly, this is particularly true for more expert players, who have a higher chance of sub-optimal encounters in opening, possibly because of the depth of responses and variations within opening lines coming from a skilled opponent.

The study we have presented has two limitations. First, we kept our focus on openings only, which constitute one of the many phases of chess games. Nevertheless, this simple approach proved to be enough to reveal how experts differ from beginners in simple quantifiable ways. It also complements existing work on recall abilities of players for chess positions as a function of skill level~\cite{simon1988skill}. A first natural extension in this direction would consist in analysing also other parts of the game, such as middle game and endings. A second limitation is that, when associating a skill level to a player, we inevitably considered Elo as a static, immutable measure. Instead, this rating systems is clearly in constant evolution throughout the career of a player. While including this dynamical aspect of ranking would surely add a missing aspect to the analysis, it is worth stressing that our measure is still a good proxy for skill level, as we have neglected the initial phase of the careers---associated to the steepest growth/change in Elo.

Taken together, our work represents a first step towards understanding the game mechanisms associated to performance in the careers of chess players. Future work might enrich this analysis by considering the complexity of chess games as a whole via considering the full sequences of moves instead of focusing on the important phases of the game only.  
Finally, and more broadly, it would be interesting to extend our approach to other ecosystems, investigating tensions between specialization and diversity in other contexts---from Go to tennis and boxing matches---where opening and response constitute a crucial part of the game.

\section{Materials and Methods}
\subsection{Data}
\label{data}
For our analysis, we use all games played on the online chess server~\emph{lichess.org} between 2013 and 2016. Such data covers 123 million games played between 0.98 million players. There are different games available to the players on the platform: bullet, blitz, and rapid. The analysis presented in this paper is restricted to Blitz games, which are fast and tactical but still allow for some strategy in the game overall unlike bullet games which last only 1 minute at most and littered with pre-moves. The most popular time controls for blitz are 5 mins and 3 mins. We specifically focused on this type of games since ``speed'' chess is played across all levels, from beginners to grandmasters. 

\subsection{Measuring Diversity}
We measure the diversity of openings of a player by calculating the Shannon entropy~\cite{shannon2001mathematical} of the distribution of frequency of opening moves or opening sequences (see Fig. \ref{fig2} and Fig. \ref{fig3} respectively). Notice that for the analysis in Fig.\ref{fig2}b, we selected only games where the player starts as white.

\subsection{Null models for hot and cold streaks} \label{null_hot}
To calculate the expected lengths of hot streaks in a player's career, we build a null model where we reshuffle the temporal order of the associated sequence of games, thus preserving the total number of victories, losses and draws. Then, we compute the length of each hot and cold streak (sets of consecutive wins and loses) observed in this reshuffled sequence. The presence of hot- and cold-streak phenomena can be then investigated by comparing the number of hot and cold streaks of a given length $\ell$ in the actual careers with respect to these reshuffled sequences.

\subsection{Chess concepts}
\subsubsection{Openings} \label{opening}
{A chess opening is the initial stage of a chess game---a sequence of first few moves. It usually consists of established theory; the other phases are the middlegame and the endgame. All games can be associated with a unique main opening line, within which there can be many variations. Many opening sequences have standard names such as the Sicilian Defense, Ruy Lopez, Italian Game, Scotch Game etc.}

\subsubsection{Elo rating} \label{elo} {The \emph{Elo} rating system is a method for calculating the relative skill levels of players in zero-sum games such as chess. A player's Elo rating is represented by a number which may change depending on the outcome of rated games played. 

We show the Elo ratings of all players in our dataset in Supplementary Information (Sec.~S1). The career lengths of players as a function of their skill is shown in Supplementary Information (Sec.~S2). Experts tend to play more than beginners.

After every game, the winning player takes points from the losing one. The difference between the ratings of the winner and loser determines the total number of points gained or lost after a game. Two players with equal ratings who play against each other are expected to score an equal number of wins. A player whose rating is 100 points greater than their opponent's is expected to score 64 $\%$. For each player, we work with the \emph{Elo} averaged in all their games (expect the first 100 games when it fluctuates a lot).} We note that players of similar skill levels (Elo ratings) are matched to compete.

\subsubsection{Separating players by skill level}{We separated the player into the 4 skill levels as follows. We first arranged the players in ascending order of their \emph{Elo} rating (average calculated over all their games). We then created Elo bins that divide players in 4 equally sized skill categories. Finally, we labelled these bins as---\emph{ beginner, intermediate, advanced, expert} respectively.}
\subsubsection{Opening variations}{For a given chess opening there are multiple variations, as players can explore different moves after the main opening line is established. For example, the \textit{Sicilian Defence} begins with the following moves 1. e4 c5. The \textit{Sicilian Defence: Najdorf Variation} of Sicilian is 1.e4 c5 2.Nf3 d6 3.d4 cxd4 4.Nxd4 Nf6 5.Nc3 a6, while the \textit{Sicilian Defence: Dragon Variation} is 1.e4 c5 2.Nf3 d6 3.d4 cxd4 4.Nxd4 Nf6 5.Nc3 g6.}

\section*{Data availability}
The data from \href{lichess.org}{lichess.org} used in this work is openly accessible for download from \href{https://database.lichess.org/}{https://database.lichess.org/}.

\section*{Code availability}
The code used in this study is available at \href{https://github.com/chowdhary-sandeep/lichess.git}{https://github.com/chowdhary-sandeep/lichess.git}.


\bibliographystyle{jabbrv_unsrt} 

\bibliography{main}

\begin{thebibliography}{10}

\bibitem{sinatra2016quantifying}
Roberta Sinatra, Dashun Wang, Pierre Deville, Chaoming Song, and
  Albert-L{\'a}szl{\'o} Barab{\'a}si.
\newblock Quantifying the evolution of individual scientific impact.
\newblock {\em\JournalTitle{Science}}, 354(6312), 2016.

\bibitem{deville2014career}
Pierre Deville, Dashun Wang, Roberta Sinatra, Chaoming Song, Vincent~D Blondel,
  and Albert-L{\'a}szl{\'o} Barab{\'a}si.
\newblock Career on the move: Geography, stratification and scientific impact.
\newblock {\em\JournalTitle{Scientific Reports}}, 4(1):1--7, 2014.

\bibitem{jia2017quantifying}
Tao Jia, Dashun Wang, and Boleslaw~K Szymanski.
\newblock Quantifying patterns of research-interest evolution.
\newblock {\em\JournalTitle{Nature Human Behaviour}}, 1(4):1--7, 2017.

\bibitem{zeng2019increasing}
An~Zeng, Zhesi Shen, Jianlin Zhou, Ying Fan, Zengru Di, Yougui Wang, H~Eugene
  Stanley, and Shlomo Havlin.
\newblock Increasing trend of scientists to switch between topics.
\newblock {\em\JournalTitle{Nature Communications}}, 10(1):1--11, 2019.

\bibitem{fortunato2018science}
Santo Fortunato, Carl~T Bergstrom, Katy B{\"o}rner, James~A Evans, Dirk
  Helbing, Sta{\v{s}}a Milojevi{\'c}, Alexander~M Petersen, Filippo Radicchi,
  Roberta Sinatra, Brian Uzzi, et~al.
\newblock Science of science.
\newblock {\em\JournalTitle{Science}}, 359(6379), 2018.

\bibitem{liu2018hot}
Lu~Liu, Yang Wang, Roberta Sinatra, C~Lee Giles, Chaoming Song, and Dashun
  Wang.
\newblock Hot streaks in artistic, cultural, and scientific careers.
\newblock {\em\JournalTitle{Nature}}, 559(7714):396--399, 2018.

\bibitem{fraiberger2018quantifying}
Samuel~P Fraiberger, Roberta Sinatra, Magnus Resch, Christoph Riedl, and
  Albert-L{\'a}szl{\'o} Barab{\'a}si.
\newblock Quantifying reputation and success in art.
\newblock {\em\JournalTitle{Science}}, 362(6416):825--829, 2018.

\bibitem{Williams2019QuantifyingAP}
Oliver~E. Williams, Lucas Lacasa, and Vito Latora.
\newblock Quantifying and predicting success in show business.
\newblock {\em\JournalTitle{Nature Communications}}, 10, 2019.

\bibitem{Janosov2020SuccessAL}
Mil{\'a}n Janosov, Federico Battiston, and Roberta Sinatra.
\newblock Success and luck in creative careers.
\newblock {\em\JournalTitle{EPJ Data Science}}, 9:1--12, 2020.

\bibitem{Janosov2020ElitesCA}
Mil{\'a}n Janosov, Federico Musciotto, Federico Battiston, and Gerardo
  I{\~n}iguez.
\newblock Elites, communities and the limited benefits of mentorship in
  electronic music.
\newblock {\em\JournalTitle{Scientific Reports}}, 10, 2020.

\bibitem{stevenson1990early}
Christopher~L Stevenson.
\newblock The early careers of international athletes.
\newblock {\em\JournalTitle{Sociology of Sport Journal}}, 7(3):238--253, 1990.

\bibitem{conzelmann2003professional}
Achim Conzelmann and Siegfried Nagel.
\newblock Professional careers of the german olympic athletes.
\newblock {\em\JournalTitle{International review for the sociology of sport}},
  38(3):259--280, 2003.

\bibitem{mallett2004elite}
Clifford~J Mallett and Stephanie~J Hanrahan.
\newblock Elite athletes: why does the ‘fire’burn so brightly?
\newblock {\em\JournalTitle{Psychology of sport and exercise}}, 5(2):183--200,
  2004.

\bibitem{stambulova2013athletes}
Natalia~B Stambulova and Tatiana~V Ryba.
\newblock {\em Athletes' careers across cultures}, volume 288.
\newblock Routledge London, 2013.

\bibitem{lewis2004moneyball}
Michael Lewis.
\newblock {\em Moneyball: The art of winning an unfair game}.
\newblock WW Norton \& Company, 2004.

\bibitem{radicchi2011best}
Filippo Radicchi.
\newblock Who is the best player ever? a complex network analysis of the
  history of professional tennis.
\newblock {\em\JournalTitle{PloS one}}, 6(2):e17249, 2011.

\bibitem{Nevill2008TwentyfiveYO}
Alan~Michael Nevill, Greg Atkinson, and Mike Hughes.
\newblock Twenty-five years of sport performance research in the journal of
  sports sciences.
\newblock {\em\JournalTitle{Journal of Sports Sciences}}, 26:413 -- 426, 2008.

\bibitem{andrew2019research}
Damon~PS Andrew, Paul~M Pedersen, and Chad~D McEvoy.
\newblock {\em Research methods and design in sport management}.
\newblock Human Kinetics, 2019.

\bibitem{schaigorodsky2014memory}
Ana~L Schaigorodsky, Juan~I Perotti, and Orlando~V Billoni.
\newblock Memory and long-range correlations in chess games.
\newblock {\em\JournalTitle{Physica A: Statistical Mechanics and its
  Applications}}, 394:304--311, 2014.

\bibitem{blasius2009zipf}
Bernd Blasius and Ralf T{\"o}njes.
\newblock Zipf’s law in the popularity distribution of chess openings.
\newblock {\em\JournalTitle{Phys. Rev. Lett.}}, 103(21):218701, 2009.

\bibitem{arabaci2006investigation}
Ramiz Arabac{\i}.
\newblock An ınvestıgatıon ınto the openıngs used by top 100 chess
  players.
\newblock {\em\JournalTitle{International Journal of Performance Analysis in
  Sport}}, 6(1):149--160, 2006.

\bibitem{bilalic2007does}
Merim Bilali{\'c}, Peter McLeod, and Fernand Gobet.
\newblock Does chess need intelligence?—a study with young chess players.
\newblock {\em\JournalTitle{Intelligence}}, 35(5):457--470, 2007.

\bibitem{charness2005role}
Neil Charness, Michael Tuffiash, Ralf Krampe, Eyal Reingold, and Ekaterina
  Vasyukova.
\newblock The role of deliberate practice in chess expertise.
\newblock {\em\JournalTitle{Applied Cognitive Psychology}}, 19(2):151--165,
  2005.

\bibitem{campitelli2011deliberate}
Guillermo Campitelli and Fernand Gobet.
\newblock Deliberate practice: Necessary but not sufficient.
\newblock {\em\JournalTitle{Current directions in psychological science}},
  20(5):280--285, 2011.

\bibitem{hambrick2014deliberate}
David~Z Hambrick, Frederick~L Oswald, Erik~M Altmann, Elizabeth~J Meinz,
  Fernand Gobet, and Guillermo Campitelli.
\newblock Deliberate practice: Is that all it takes to become an expert?
\newblock {\em\JournalTitle{Intelligence}}, 45:34--45, 2014.

\bibitem{gilovich1985hot}
Thomas Gilovich, Robert Vallone, and Amos Tversky.
\newblock The hot hand in basketball: On the misperception of random sequences.
\newblock {\em\JournalTitle{Cognitive Psychology}}, 17(3):295--314, 1985.

\bibitem{miller2018surprised}
Joshua~B Miller and Adam Sanjurjo.
\newblock Surprised by the hot hand fallacy? a truth in the law of small
  numbers.
\newblock {\em\JournalTitle{Econometrica}}, 86(6):2019--2047, 2018.

\bibitem{liu2021understanding}
Lu~Liu, Nima Dehmamy, Jillian Chown, C~Lee Giles, and Dashun Wang.
\newblock Understanding the onset of hot streaks across artistic, cultural, and
  scientific careers.
\newblock {\em\JournalTitle{Nature Communications}}, 12(1):1--10, 2021.

\bibitem{young1999zone}
Janet~A Young and Michelle~D Pain.
\newblock The zone: Evidence of a universal phenomenon for athletes across
  sports.
\newblock {\em\JournalTitle{Athletic Insight: the online journal of sport
  psychology}}, 1(3):21--30, 1999.

\bibitem{murphy2011zone}
Michael Murphy and Rhea~A White.
\newblock {\em In the zone: Transcendent experience in sports}.
\newblock Open Road Media, 2011.

\bibitem{kuhn2011essential}
T.S. Kuhn.
\newblock {\em The Essential Tension: Selected Studies in Scientific Tradition
  and Change}.
\newblock University of Chicago Press, 2011.

\bibitem{march1991exploration}
James~G March.
\newblock Exploration and exploitation in organizational learning.
\newblock {\em\JournalTitle{Organization Science}}, 2(1):71--87, 1991.

\bibitem{pappalardo2015returners}
Luca Pappalardo, Filippo Simini, Salvatore Rinzivillo, Dino Pedreschi, Fosca
  Giannotti, and Albert-L{\'a}szl{\'o} Barab{\'a}si.
\newblock Returners and explorers dichotomy in human mobility.
\newblock {\em\JournalTitle{Nature Communications}}, 6(1):1--8, 2015.

\bibitem{iacopini2018network}
Iacopo Iacopini, Sta{\v{s}}a Milojevi{\'c}, and Vito Latora.
\newblock Network dynamics of innovation processes.
\newblock {\em\JournalTitle{Physical Review Letters}}, 120(4):048301, 2018.

\bibitem{ericsson1996expert}
K~Anders Ericsson and Andreas~C Lehmann.
\newblock Expert and exceptional performance: Evidence of maximal adaptation to
  task constraints.
\newblock {\em\JournalTitle{Annual Review of Psychology}}, 47(1):273--305,
  1996.

\bibitem{Ericsson1996TheRT}
K.~Anders Ericsson.
\newblock The road to excellence: The acquisition of expert performance in the
  arts and sciences, sports, and games.
\newblock 1996.

\bibitem{Groot1978ThoughtAC}
Adrianus~Dingeman de~Groot.
\newblock Thought and choice in chess.
\newblock 1978.

\bibitem{simon1988skill}
Herbert Simon and William Chase.
\newblock Skill in chess.
\newblock In {\em Computer chess compendium}, pages 175--188. Springer, 1988.

\bibitem{Chase1973PerceptionIC}
William~G. Chase and Herbert~A. Simon.
\newblock Perception in chess.
\newblock {\em\JournalTitle{Cogn. Psychol.}}, 4:55--81, 1973.

\bibitem{Simon1973ASO}
Herbert~A. Simon and Kevin~Michael Gilmartin.
\newblock A simulation of memory for chess positions.
\newblock {\em\JournalTitle{Cognitive Psychology}}, 5:29--46, 1973.

\bibitem{de2022quantifying}
Giordano De~Marzo and Vito~DP Servedio.
\newblock Quantifying the complexity and similarity of chess openings using
  online chess community data.
\newblock {\em\JournalTitle{arXiv preprint arXiv:2206.14312}}, 2022.

\bibitem{shannon2001mathematical}
Claude~Elwood Shannon.
\newblock A mathematical theory of communication.
\newblock {\em\JournalTitle{ACM SIGMOBILE mobile computing and communications
  review}}, 5(1):3--55, 2001.

\end{thebibliography}

\section*{Acknowledgements}
I.I. acknowledges support from the James S. McDonnell Foundation $21^{\text{st}}$ Century Science Initiative Understanding Dynamic and Multi-scale Systems - Postdoctoral Fellowship Award.

 \section*{Author contributions}
S.C. wrote the code and performed the numerical analysis. S.C., I.I., F.B. interpreted the results. S.C., I.I., F.B. wrote the manuscript.

\section*{Competing interests}
The authors declare that they have no competing interests.

\section*{Additional information}
{\bf Supplementary Information} is provided as a separate file.

\section*{Materials \& correspondence}
Correspondence should be addressed to \url{Chowdhary_Sandeep@phd.ceu.edu}.


\end{document}


\begin{center}
{\LARGE Supplementary Information for}\\[0.7cm]
{\Large \textbf{Quantifying human performance in chess}}\\[0.5cm]
{\large S. Chowdhary, I. Iacopini, F. Battiston}\\[0.7cm]
{\small $^*$Corresponding author email: \href{mailto:chowdhary_sandeep@phd.ceu.edu}{chowdhary\_sandeep@phd.ceu.edu}
}\\[2cm]
\end{center}


\section{ELO distribution of the population}
\label{sec:elo_supp}


\begin{figure}[h]
    \centering
  
    \begin{minipage}{1\textwidth}
        \centering
        \includegraphics[width=.6\textwidth]{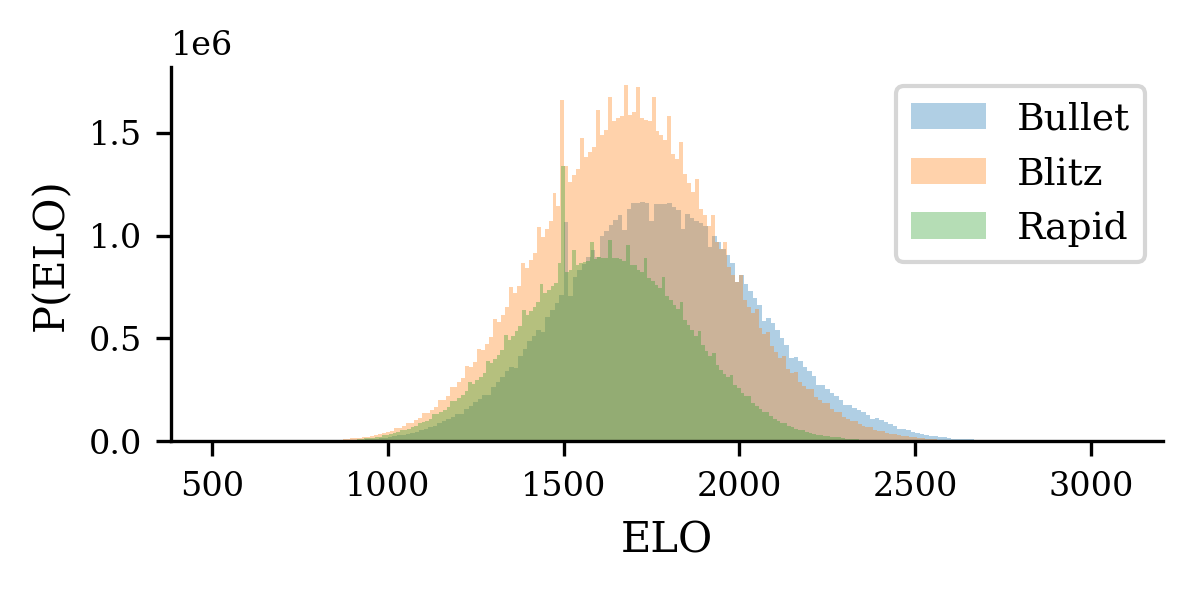} 
    \end{minipage}
    \caption{Distribution of ELO ratings averaged over the career of a player separately for the different time controls i.e. Bullet, Blitz and Rapid.}
    \label{fig:fig_filtering}
\end{figure}

\section{Career length distributions vs skill level}
\label{sec:career_lengths_supp}


\begin{figure}[h]
    \centering
  
    \begin{minipage}{1\textwidth}
        \centering
        \includegraphics[width=1\textwidth]{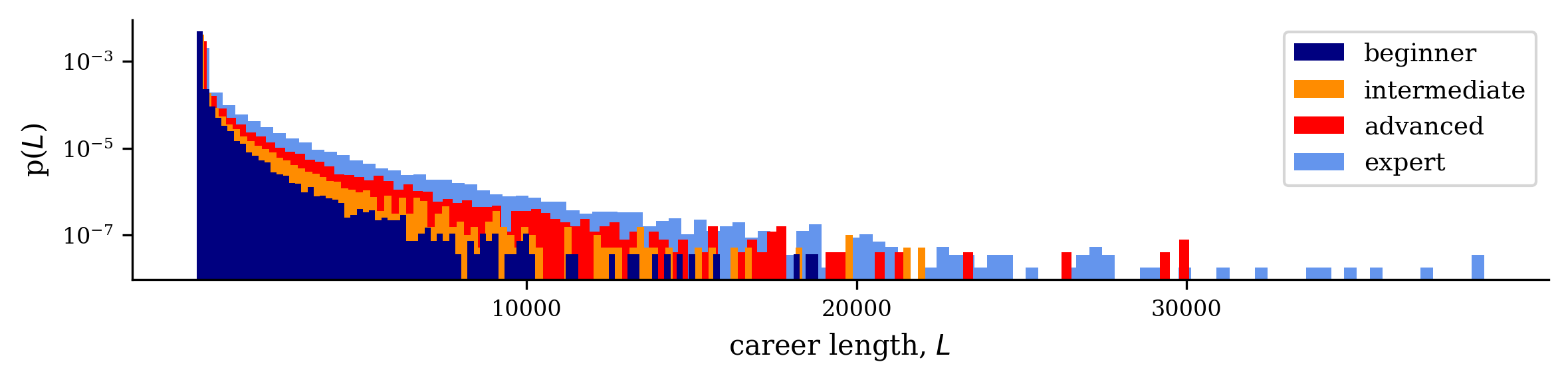} 
    \end{minipage}
    \caption{Distribution of the total number of games by a player for the 4 skill categories}
    \label{fig:fig_filtering}
\end{figure}
\clearpage

\section{Opening switches in a career vs skill level}
\label{sec:opening_supp}
We calculate the number of opening switches between consecutive games for each player in the dataset, and compare them with the number expected in a reshuffled null model for each player. For the null model, we reshuffle the temporal order of the associated sequence of games, thus preserving the total number of victories, losses and draws.

\begin{figure}[h]
    \centering
  
    \begin{minipage}{1\textwidth}
        \centering
        \includegraphics[width=.6\textwidth]{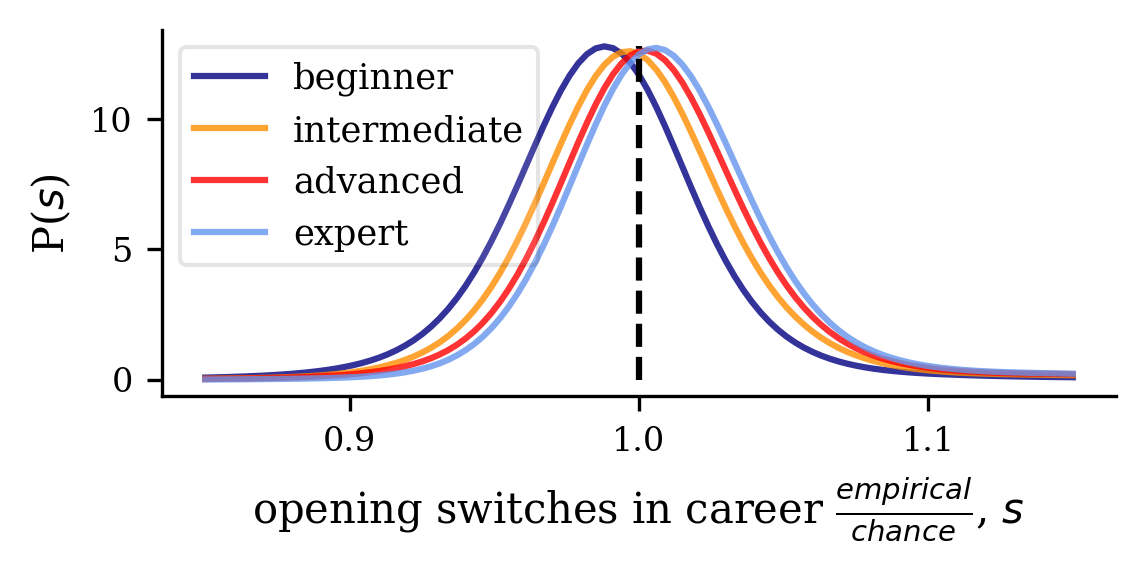} 
    \end{minipage}
    \caption{Distribution of the number of opening switches in a player's career properly normalised with the null model aggregated into 4 skill categories.}
    \label{fig:fig_filtering}
\end{figure}



  
